\documentclass[letterpaper]{article} 
\usepackage{aaai25}  
\usepackage{times}  
\usepackage{helvet}  
\usepackage{courier}  
\usepackage[hyphens]{url}  
\usepackage{graphicx} 
\usepackage{natbib}  
\usepackage{caption} 
\usepackage{amsmath,amssymb,amsfonts}
\usepackage{multirow}
\usepackage{booktabs}
\usepackage{algorithm}
\usepackage{algorithmic}
\usepackage{cases}

\usepackage{newfloat}
\usepackage{listings}
\DeclareCaptionStyle{ruled}{labelfont=normalfont,labelsep=colon,strut=off} 
\floatstyle{ruled}
\newfloat{listing}{tb}{lst}{}
\floatname{listing}{Listing}
\title{Unsupervised Self-Prior Embedding Neural Representation for Iterative Sparse-View CT Reconstruction}
\author {
    Xuanyu Tian\textsuperscript{\rm 1,\rm 2},
    Lixuan Chen\textsuperscript{\rm 3},
    Qing Wu\textsuperscript{\rm 1},
    Chenhe Du\textsuperscript{\rm 1},
    Jingjing Shi\textsuperscript{\rm 4},
    Hongjiang Wei\textsuperscript{\rm 5}, \\
    Yuyao Zhang\textsuperscript{\rm 1}\thanks{Corresponding author.} 
}
\affiliations {
    \textsuperscript{\rm 1} School of Information Science and Technology, ShanghaiTech University, Shanghai 201210, China\\
    \textsuperscript{\rm 2} Lingang Laboratory, Shanghai 200031, China\\
    \textsuperscript{\rm 3} Electrical and Computer Engineering, University of Michigan, MI 48105, United States\\
    \textsuperscript{\rm 4} Tongji Hospital, Tongji Medical College, Huazhong University of Science and Technology, Wuhan 430030, China\\
    \textsuperscript{\rm 5}School of Biomedical Engineering, Shanghai Jiao Tong University, Shanghai 200127, China\\
    tianxy@shanghaitech.edu.cn, chenlx@umich.edu, wuqing@shanghaitech.edu.cn, duchenhe@shanghaitech.edu.cn, walm13@126.com, hongjiang.wei@sjtu.edu.cn, zhangyy8@shanghaitech.edu.cn
}

\usepackage{bibentry}

\begin{document}

\maketitle

\begin{abstract}
Emerging unsupervised implicit neural representation (INR) methods, such as NeRP, NeAT, and SCOPE, have shown great potential to address sparse-view computed tomography (SVCT) inverse problems. Although these INR-based methods perform well in relatively dense SVCT reconstructions, they struggle to achieve comparable performance to supervised methods in sparser SVCT scenarios. They are prone to being affected by noise, limiting their applicability in real clinical settings. Additionally, current methods have not fully explored the use of image domain priors for solving SVCsT inverse problems. In this work, we demonstrate that imperfect reconstruction results can provide effective image domain priors for INRs to enhance performance. To leverage this, we introduce \textbf{S}elf-\textbf{p}rior \textbf{e}mbedding \textbf{ne}ural \textbf{r}epresentation (\textbf{Spener}), a novel unsupervised method for SVCT reconstruction that integrates iterative reconstruction algorithms. During each iteration, Spener extracts local image prior features from the previous iteration and embeds them to constrain the solution space. Experimental results on multiple CT datasets show that our unsupervised Spener method achieves performance comparable to supervised state-of-the-art (SOTA) methods on in-domain data while outperforming them on out-of-domain datasets. Moreover, Spener significantly improves the performance of INR-based methods in handling SVCT with noisy sinograms.
Our code is available at \url{https://github.com/MeijiTian/Spener}.
\end{abstract}

%

\section{Introduction}
X-ray computed tomography (CT) is widely used in clinical diagnosis due to its non-invasive nature and high efficiency~\cite{wang2008outlook}. 
However, the ionizing radiation of X-ray poses potential risks to biological tissues, while high levels of radiation exposure can increase the lifetime risk of cancer.
To mitigate these risks, the medical community generally follows the ALARA (as low as reasonably achievable) principle in clinical practice. 
Therefore, reducing the radiation dose in the CT acquisition process is a critical and widely researched topic.

The forward acquisition of CT can be formulated as the following linear equation:
\begin{equation}
    \mathbf{y} = \mathbf{A}\mathbf{x} + \boldsymbol{\epsilon},
\label{eq:forward}
\end{equation}
where the $\mathbf{y} \in \mathbb{R}^{m}$ is the measurement data (\textit{i.e.} sinogram data) with $m = n_v \times n_d$, $n_v$ and $n_d$ is the number of projection views and detectors. $\mathbf{A}\in \mathbb{R}^{m\times n}$ is the forward model (\textit{e.g.} Radon transform), $\mathbf{x} \in \mathbb{R}^{n}$ is the unknown image intensity of object and $\boldsymbol{\epsilon} \in \mathbb{R}^{m}$ is the system noise.
To reduce the radiation of the CT process, one effective solution is to reduce the number of sampling projection views $n_v$, which is sparse-view CT (SVCT) acquisition. 
Under the SVCT acquisition setting, the dimension of measurement data $\mathbf{y}$ is much lower than the unknown object $\mathbf{x}$ (\textit{i.e}, $m \ll n$), resulting in the linear equation (Eq.~\eqref{eq:forward}) to be highly ill-posed. 



Recently, there has been considerable works on supervised deep learning (DL) for the SVCT inverse problem~\cite{lee2018deep,zhang2018sparse,chen2018learn,wu2021drone,ma2023freeseed, RegFormer}.  
These methods train neural networks to map corrupted images to artifact-free images in end-to-end learning.
Leveraging the power of neural networks and data-driven priors, supervised DL methods achieve SOTA performance in SVCT reconstruction. 
However, these methods face two significant challenges:
1) Learning cost. Developing effective data-driven priors requires collecting large amounts of paired data, which is both challenging and expensive;
2) Out-of-domain (OOD) problem. These methods often suffer from performance degradation when the CT acquisition settings change (\textit{e.g.}, different organs and X-ray acquisition geometry) deviate from the training data.
These issues pose substantial obstacles to the practical application of supervised methods in clinical settings.

Implicit Neural Representation (INR) has recently become an innovative unsupervised approach for solving imaging inverse problems without relying on external data.
It utilizes multi-layer perception (MLP) to represent the intensity of an object into an implicit function of spatial coordinates. 
By integrating differentiable physical models, the unknown image intensity can be transformed into the measurement domain. 
Thus, MLP can learn the implicit function by minimizing the error between the estimated measurements and the acquired data.  
The inherent learning consistency bias of INR acts as an effective implicit regularization term for solving inverse problems~\cite{rahaman2019spectral},
achieving significant progress in SVCT reconstruction~\cite{sun2021coil, zang2021intratomo, reed2021dynamic, zha2022naf, wu2022self, lin2023learning,lin2024c}.
However, existing INR-based SVCT methods rely solely on the implicit learning bias of the network as a regularization constraint. When the ill-posedness increases in SVCT reconstruction (\textit{e.g.} sparser view or lower dose in acquisition), they exhibit instability to provide satisfactory performance. 


In this work, we demonstrate that the imperfectly reconstructed result of INR can be an effective image domain prior to stabilizing the neural representation. 
Therefore, we propose a novel unsupervised method, \textbf{S}elf-\textbf{p}rior \textbf{e}mbedding \textbf{ne}ural \textbf{r}epresentation (\textbf{Spener}), which explores the image domain prior generated by INR itself for SVCT reconstruction.
To utilize the local image features in the image domain, we incorporate an image encoder to represent the prior image into local image feature vectors providing INR network optimization.
Additionally, we employ an iterative reconstruction algorithm, such as plug-and-play half-quadratic splitting (PnP-HQS), to iteratively update the prior image to accelerate the convergence of INR and enhance its performance, particularly in highly ill-posed scenarios.

We conduct extensive experiments to evaluate the performance of our Spener on three public datasets. The results show that Spener achieves comparative performance with the supervised SOTA methods in the in-distribution datasets while outperforming them when the CT acquisition parameters are different from those in the training data. 
Meanwhile, in noisy acquisition scenarios, Spener significantly outperforms other unsupervised INR methods, highlighting its superiority.
Extensive ablation studies further validate the effectiveness and robustness of the proposed method. 
Our main contributions are summarized as follows:
\begin{itemize}
    \item We propose Spener, a novel unsupervised robust neural representation, achieving superior performance in solving SVCT reconstruction.
    \item We demonstrate that imperfect INR reconstruction can serve as an effective image prior and explicit regularization constraint for INR optimization. 
    \item We introduce an image encoder to provide the INR network with local continuous image priors, with experimental results confirming its effectiveness in unsupervised learning. 
    \item We integrate the strengths of iterative algorithms with INR-based methods, improving the model convergence and generalization. 
\end{itemize}

\section{Related Work}

\subsection{Implicit Neural Representation for CT}
Consider a 2D CT scanner, the image intensity of scanned object can be formulated as a function of spatial coordinates, defined by
\begin{equation}
    f: \mathbf{p} = (x,y) \in \mathbb{R}^2 \longrightarrow \mu(\mathbf{p}) \in \mathbb{R},
\label{eq:inr_ct}
\end{equation}
where $\mathbf{p}$ is an arbitrary coordinate and $\mu(\mathbf{p})$ is the corresponding intensity of the object. However, the function $f$ is highly complex, making an analytical solution intractable. Previous works~\cite{shen2022nerp,zha2022naf,wu2022self} have leveraged the consistency bias inherent in INR by employing an MLP network $\mathcal{F}_{\mathbf{\Phi}}$ to approximate the function $f$.
The MLP network can learn the implicit function by incorporating the forward model (\textit{e.g.} Radon transform) to minimize the discrepancy between the predicted and acquired measurement data:
\begin{equation}
    \mathbf{\Phi}^* = \underset{\mathbf{\Phi}}{\arg\min} \mathcal{L}(\mathbf{A}\mathcal{F}_\mathbf{\Phi}, \mathbf{y}),
\end{equation}
where $\mathbf{A}$ is the forward model and $\mathbf{y}$ is the acquired measurement data.
By adopting the novel coordinate encoding strategy~\cite{muller2022instant}, these INR-based methods can effectively capture high-frequency signals, providing high-quality reconstructions while maintaining data consistency. However, when the measurement data is undersampled or noisy, these methods tend to generate high-frequency artifacts, resulting in performance degradation.
\begin{figure*}[!t]
\centering
\includegraphics[width=\textwidth]{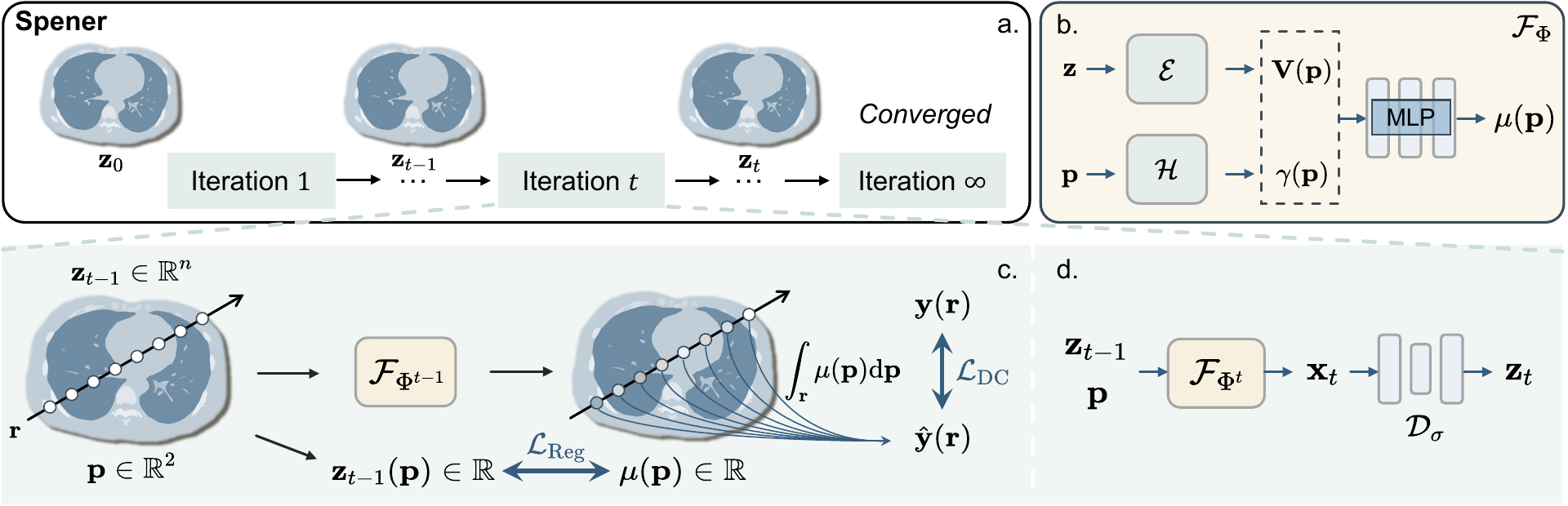} 
\caption{\textbf{Overview of Spener model}, including (a) iterative reconstruction using an image embedding neural network $\mathcal{F}_\Phi$, (b) architecture of the image embedding neural network $\mathcal{F}_\Phi$, (c) solving the data fidelity subproblem via the image embedding neural network $\mathcal{F}_\Phi$, and (d) solving regularization subproblem via a denoiser $\mathcal{D}_\sigma$.}
\label{fig:pipeline}
\end{figure*}

\subsection{Plug-and-Play Half-Quadratic Splitting}
The classical approach for solving the ill-posed problem in Eq.~(\ref{eq:forward}) is by formulating an optimization problem:
\begin{equation}
    \hat{\mathbf{x}} = \underset{\mathbf{x}}{\arg \min}\frac{1}{2} \| \mathbf{y} -\mathbf{A} \mathbf{x}\|_2^2 + \lambda \mathcal{R}(\mathbf{x}),
\label{eq1}
\end{equation}
where the objective function is composed of a data fidelity term $ \| \mathbf{y} -\mathbf{A} \mathbf{x}\|_2^2/2$ and a regularization term $\lambda \mathcal{R}(\mathbf{x})$. 
The data fidelity term ensures that the solution conforms to the forward process, maintaining consistency with the measurement data, while the regularization enforces prior knowledge, narrowing the solution space. 

Plug-and-play half-quadratic splitting (PnP-HQS) is a widely used iterative framework for solving inverse problems~\cite{zhang2021plug}. 
The conventional HQS~\cite{geman1995nonlinear} introduces an auxiliary variable $\mathbf{z}$ to reformulate the original optimization problem Eq.~(\ref{eq1}) in a constrained manner:
\begin{equation}
    \hat{\mathbf{x}} = \underset{\mathbf{x}}{\arg\min}\frac{1}{2}\|\mathbf{y} - \mathbf{A}{\mathbf{x}} \|^2 + \lambda\mathcal{R}(\mathbf{z})\quad \textit{s.t.}\quad \mathbf{z} = \mathbf{x}.
\end{equation}
HQS solves the objective function by decoupling a data fidelity term and a regularization term, and alternatively solving them in iterative ways:
\begin{subequations}
    \label{eq:m1.hqs}
    \begin{numcases}{}
        \mathbf{x}_{t}=\underset{\mathbf{x}}{\mathrm{arg} \min}\left\| \mathbf{y}-\mathbf{A}\mathbf{x} \right\| ^2+\frac{\mu}{2} \left\| \mathbf{x}-\mathbf{z}_{t-1} \right\| ^2\label{eq:m1.data_sub}\\
        \mathbf{z}_{t}=\underset{\mathbf{z}}{\mathrm{arg} \min}\frac{1}{2(\sqrt{\lambda /\mu})^2}\left\| \mathbf{z}-\mathbf{x}_{t} \right\| ^2+\mathcal{R} \left( \mathbf{z} \right)\label{eq:m1.prior_sub}.
    \end{numcases}
\end{subequations}
The PnP-HQS proposes utilizing a general denoising algorithm to provide the solution of Eq~(\ref{eq:m1.prior_sub}), given by
\begin{equation}
    \mathbf{z}_t = \mathcal{D}_{\sigma}(\mathbf{x}_{t}),
    \label{eq:pnp_reg}
\end{equation}
where $\mathcal{D}_{\sigma}$ represents the denoising operation with a noise level parameter 
$\sigma$.

In Spener, we introduce an INR-based approach to solve the data fidelity sub-problem.
Subsequently, we apply an effective denoiser to regularize the reconstruction obtained from the INR. By iteratively solving these two sub-problems, we achieve a high-fidelity reconstruction.

\section{Proposed Method}
\subsection{Overview Spener}
Figure~\ref{fig:pipeline} shows the overall pipeline of Spener.
(a) Starting with an initial result $\mathbf{z}_0$, Spener iteratively refines the CT image using the image domain embedding neural network $\mathcal{F}_\mathbf{{\Phi}}$.
(b) The network $\mathcal{F}_\mathbf{{\Phi}}$ leverage the prior image $\mathbf{z}$ and spatial coordinates $\mathbf{p}$ to represent the underlying CT image intensity $\mu(\mathbf{p})$.
To further exploit the local image domain prior, $\mathcal{F}_\mathbf{{\Phi}}$ incorporates an image encoder $\mathcal{E}$ to extract the local image features. 
(c) At each iteration $t$, the INR-based network addresses the data fidelity subproblem. By incorporating the CT physical forward model, the network output $\mu(\mathbf{p})$ is transformed into the measurement data $\mathbf{\hat{y}}(\mathbf{r})$. Finally, the network $\mathcal{F}_{\mathbf{\Phi}^{t-1}}$ is optimized by minimizing data consistency loss $\mathcal{L}_\text{DC}$ and regularization loss $\mathcal{L}_\text{Reg}$. 
(d) After updating the network, we can generate a new reconstruction of the CT image $\mathbf{x}_t$ from $\mathcal{F}_{\mathbf{\Phi}^{t}}$. The regularization subproblem is then addressed using an effective denoiser $\mathcal{D}_\sigma$, which provides a prior image for the next iteration. 

\subsection{Image Prior Embedding Neural Representation}
To incorporate the image domain priors, we represent the underlying clean CT images as the function of the spatial coordinates and the prior image:
\begin{equation}
    \mathcal{F}: \mathbf{z}\in\mathbb{R}^{n}, \mathbf{p}\in \mathbb{R}^2 \longrightarrow \mu(\mathbf{p}) \in \mathbb{R},
\end{equation}
where $\mathbf{z}$ represents the prior image, $\mathbf{p}$ denotes an arbitrary spatial coordinate and $\mu(\mathbf{p})$ is the image intensity of the desired CT image.
The implicit function $\mathcal{F}$ is learned by a neural network $\mathcal{F}_\mathbf{\Phi}$. 
Figure~\ref{fig:pipeline}.(b) shows the architecture of the network $\mathcal{F}_\mathbf{\Phi}$, which consists of an image encoder $\mathcal{E}$ and a coordinate encoder $\mathcal{H}$.
The image encoder $\mathcal{E}$ processes the prior image to extract local neighboring features around the coordinate $\mathbf{p}$, producing the corresponding image local feature $\mathbf{V}(\mathbf{p})$.
For the coordinate encoder $\mathcal{H}$, we use hash encoding~\cite{muller2022instant} to map the coordinates into a high-dimensional feature vector $\gamma(\mathbf{p})$, facilitating the representation of high-frequency signals.
Finally, an MLP network takes the image feature $\mathbf{V}(\mathbf{p})$ and the coordinate feature $\gamma(\mathbf{p})$ as inputs to estimate the image intensity $\mu(\mathbf{p})$.

\subsection{Network Optimization for Solving Data Fidelity}
Figure~\ref{fig:pipeline}.(c) illustrates the workflow for addressing the data fidelity term in Eq.~(\ref{eq:m1.data_sub}) using self-prior embedding neural representation. 
Leveraging the strengths of the INR, the neural network $\mathcal{F}_\mathbf{\Phi}$ can model a continuous CT image.
This allows us to apply the CT forward physical model to transform the neural representation into the measurement domain.
For clarity, we denote the acquired sinogram data $\mathbf{y}(\mathbf{r})$ as the integral intensity along an X-ray trajectory $\mathbf{r}$.
Given a sinogram data $\mathbf{y}(\mathbf{r})$, we first construct a ray $\mathbf{r}$ and uniformly sample a set of coordinates $\mathbf{p}$ with a fixed interval $\Delta \mathbf{p}$ along the ray. 
The network $\mathcal{F}_\mathbf{\Phi}$ takes the prior image $\mathbf{z}_{t-1}$ and the set of coordinates $\mathbf{p}$ as input to estimate the corresponding image intensity $\mu(\mathbf{p})$.
Finally, we apply a differential integral operation to obtain the predicted sinogram data $\hat{\mathbf{y}}(\mathbf{r})$ as:
\begin{equation}
    \hat{\mathbf{y}}(\mathbf{r}) = \sum_{\mathbf{p}\in \mathbf{r}} \mathcal{F}_{\mathbf{\Phi}^{t-1}}(\mathbf{z}_{t-1}, \mathbf{p}) \cdot \Delta \mathbf{p}.
\end{equation}
Thus, the neural network can achieve data fidelity aligned with measurement data by minimizing the following data consistency loss:
\begin{equation}
    \mathcal{L}_\text{DC} =  \frac{1}{|\mathcal{R}|} \sum_{\mathbf{r} \in \mathcal{R}} \left | \mathbf{y}(\mathbf{r}) - \hat{\mathbf{y}}(\mathbf{r})\right |.
\end{equation}
However, directly minimizing the data consistency loss may lead to overfitting.
The latter term in the data fidelity sub-problem can be formulated as the regularization loss:
\begin{equation}
    \mathcal{L}_\text{Reg} = \frac{1}{N}\sum_{i=1}^{N}\left \|\mathcal{F}_{\mathbf{\Phi}^{t-1}} (\mathbf{z}_{t-1}(\mathbf{p}_i),\mathbf{p}_i ) - \mathbf{z}_{t-1}(\mathbf{p}_i)\right \| ^2
\end{equation}
and the network parameters $\mathbf{\Phi}^{t}$ can be optimized by solving the following optimization:
\begin{equation}
  \mathbf{\Phi}^{t} = \underset{\mathbf{\Phi}^{t-1}}{\arg\min}\ \mathcal{L}_\text{DC} + \lambda \cdot \mathcal{L}_\text{Reg},
  \label{eq:loss_3}
\end{equation}
where $\lambda$ is a hyperparameter that balances the contributions of the two terms in the total loss function. Its impact is analyzed in the following ablation studies.

\subsection{Prior Image Update for Regularization Solving}
Figure~\ref{fig:pipeline}.(d) illustrates the process for solving the regularization subproblem in Eq.~(\ref{eq:m1.prior_sub}).
After the data fidelity subproblem is addressed at iteration $t$, we can generate a new reconstructed CT image $\mathbf{x}_t$ from updated network $\mathcal{F}_{\mathbf{\Phi}^{t}}$.
Specifically, $\mathcal{F}_{\mathbf{\Phi}^{t}}$ takes the prior image from the previous iteration $\mathbf{z}_{t-1}$ and the grid coordinates as input to produce $\mathbf{x}_{t}$.
Then, we leverage an effective denoiser $\mathcal{D}_{\sigma} $ to address the regularization subproblem. Formally, this process can be expressed as below:
\begin{equation}
    \mathbf{z}_t = \mathcal{D}_{\sigma}(\mathbf{x}_t),\quad \mathbf{x}_t(\mathbf{p}) = \mathcal{F}_{\mathbf{\Phi}^{t}}(\mathbf{z}_{t-1}, \mathbf{p}),
\end{equation}
where $\mathbf{z}_t$ is the solution to the regularization subproblem, serving as the new prior image for the next iteration. Numerous studies~\cite{zhang2017learning,zhang2021plug,kamilov2023plug} have shown that advanced denoisers are crucial for solving the regularization subproblem. In our Spener model, we use the classical BM3D algorithm~\cite{dabov2007image} as the denoiser. The effectiveness of denoiser regularization is further analyzed in ablation studies.

\begin{figure*}[!t]
\centering
\includegraphics[width=\textwidth]{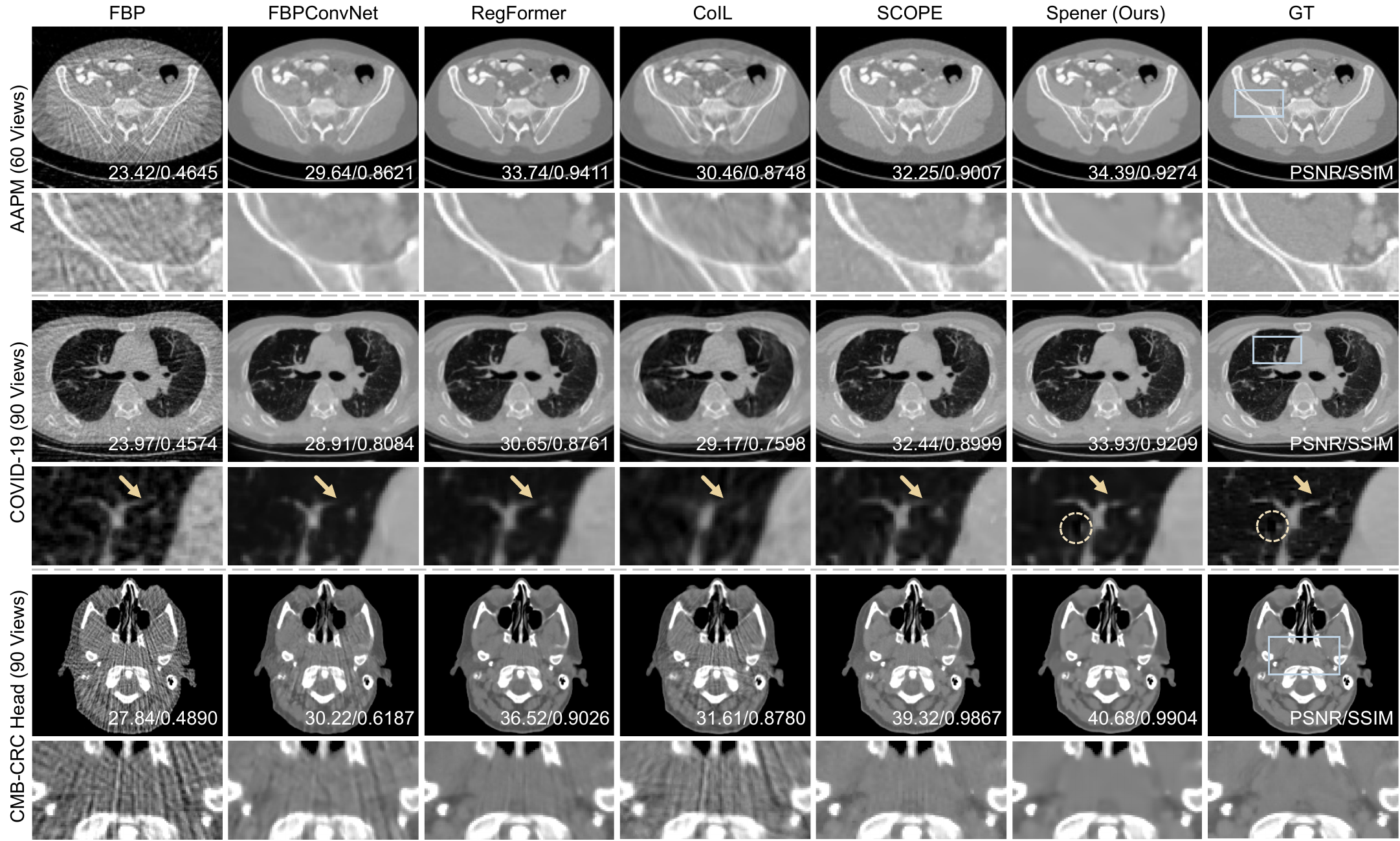} 
\caption{Qualitative results of CT images reconstructed by the compared methods on three datasets. The top two rows show results from the AAPM dataset with 60 views, the middle two rows show results from the COVID-19 dataset with 90 views, and the bottom two rows show results from the CMB-CRC head dataset with 90 views.
}
\label{fig1}
\end{figure*}

\begin{table*}[!h]
\setlength{\tabcolsep}{1.3mm}
\centering

\begin{tabular}{clcccccc} 
\toprule
\multirow{2.5}{*}{\textbf{Category}} & \multicolumn{1}{c}{\multirow{2.5}{*}{\textbf{Method}}} & \multicolumn{2}{c}{\textbf{AAPM}}                             & \multicolumn{2}{c}{\textbf{COVID-19}}                           & \multicolumn{2}{c}{\textbf{CMB-CRC Head}}                        \\ 
\cmidrule(lr){3-4}\cmidrule(lr){5-6}\cmidrule(lr){7-8}
                              & \multicolumn{1}{c}{}                & \textbf{60 Views}             & \textbf{90 Views}             & \textbf{60 Views}              & \textbf{90 Views}              & \textbf{\textbf{60 Views}} & \textbf{\textbf{90 Views}}  \\ 
\midrule
\texttt{Analytical}                    & FBP                                 & 23.61/0.4786                  & 26.46/0.6050                  & 22.96/0.4578                   & 25.77/0.5574                   & 24.05/0.4116               & 27.37/0.5086                \\ 
\cmidrule(l){1-5}\cmidrule{6-8}
\multirow{2}{*}{\texttt{Supervised}}   & FBPConvNet                          & 29.82/0.8588                  & 31.05/0.8864                  & 28.20/0.7717                   & 29.27/0.8080                   & 27.22/0.5415               & 29.74/0.6324                \\
                              & RegFormer                           & \underline{33.78}/\textbf{0.9399} & \underline{34.61}/\textbf{0.9564} & \underline{30.25}/0.8436           & 31.02/0.8810                   & 34.15/0.9264               & 36.11/0.9102                \\ 
\cmidrule(l){1-5}\cmidrule{6-8}
\multirow{3}{*}{\texttt{Unsupervised}} & CoIL                                & 30.18/0.8604                  & 31.84/0.9032                  & 27.77/0.7156                   & 29.72/0.7819                   &      28.14/0.7867                  & 31.01/0.8763                           \\
                              & SCOPE                               & 32.40/0.8939                  & 34.31/0.9322                  & 30.21/\underline{0.8504}           & \underline{32.63}/\underline{0.8997}   & \underline{36.76}/\underline{0.9740}               & \underline{38.76}/\underline{0.9857}                \\
                              & Spener (Ours)                       & \textbf{34.47}/\underline{0.9163} & \textbf{37.16}/\underline{0.9430} & \textbf{31.29}/\textbf{0.8709} & \textbf{33.73}/\textbf{0.9181} & \textbf{37.56}/\textbf{0.9809}               & \textbf{40.27}/\textbf{0.9900}                \\
\bottomrule
\end{tabular}
\caption{Quantitative results of the compared methods on AAPM, COVID-19 and CMB-CRC Head datasets. The best and second performances are highlighted in \textbf{bold} and \underline{underline}, respectively.}
\label{table1}
\end{table*}


\section{Experiments}
\subsection{Experimental Settings}
\subsubsection{Datasets} We evaluate the proposed method on three public datasets: AAPM 2016 low-dose CT grand challenge~\cite{mccollough2017low}, COVID-19 ~\cite{shakouri2021covid19} and CMB-CRC head dataset~\cite{CT-head}.

The AAPM dataset comprises 5936 full-dose CT images from 10 patients with 1mm slice thickness.  
In the experiments, we specifically select 1600 slices from 8 patients as the training set, 200 slices from 1 patient as the validation set and 10 slices from 1 patient as the test set.
\textit{The training and validation sets are prepared for the two supervised compared methods} (FBPConvNet~\cite{jin2017deep} and RegFormer~\cite{xia2022transformer}), while other compared methods (FBP~\cite{fbp}, CoIL~\cite{sun2021coil}, SCOPE~\cite{wu2022self}) and proposed Spener can directly recover high-quality image from sparse-view sinogram. 

The COVID-19 dataset comprises 3D CT volumes from over 1000 patients with confirmed COVID-19 infections. For our experiments, we chose 10 slices in the axial view as the external test data to evaluate the performance of the compared methods.
The CMB-CRC head dataset~\cite{CT-head} includes one patient head CT image with 100 slices. For our experiments, we chose 10 slices as external test data to evaluate the performance of compared methods across different organs.

\subsubsection{Dataset Simulation}
In the experiments, we retrospectively apply Radon transform on the 2D CT slice data. 
Specifically, we use \texttt{torch-radon} library~\cite{ronchetti2020torchradon} in Python with 2D fan-beam geometry to simulate the CT forward acquisition process.  
\textit{Note that we adopt the same geometry for the AAPM and CMB-CRC head datasets and other different geometry for COVID geometry.} Please refer to the supplementary materials for detailed information on the fan-beam geometry settings.

\begin{figure*}[!t]
\centering
\includegraphics[width=\textwidth]{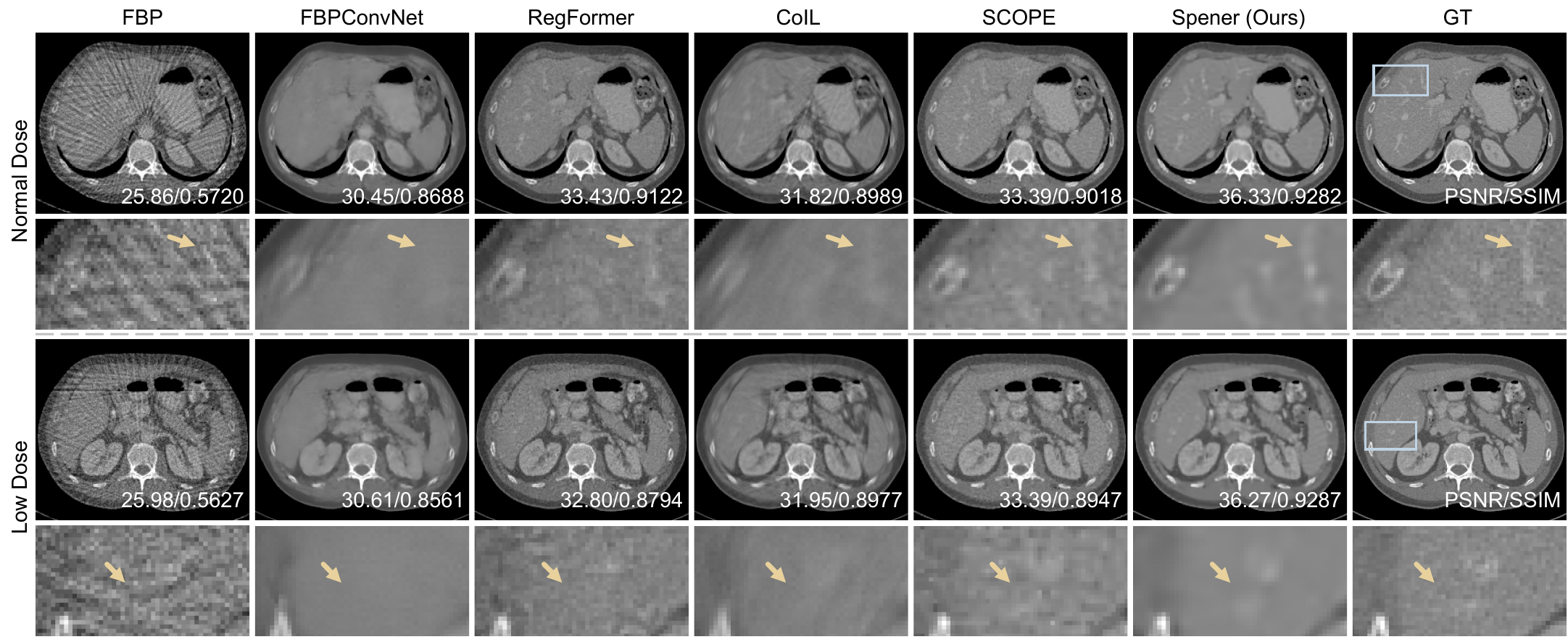} 
\caption{Qualitative results of CT image reconstructed by the compared methods under two dose settings, with both results reconstructed from AAPM dataset with 90 views.}
\label{fig2}
\end{figure*}

\begin{table*}
\centering

\begin{tabular}{clcccc} 
\toprule
\multirow{2.5}{*}{\textbf{Category}} & \multicolumn{1}{c}{\multirow{2.5}{*}{\textbf{Method}}} & \multicolumn{2}{c}{\textbf{Normal Dose} $(I_0 = 10^6)$} & \multicolumn{2}{c}{\textbf{Low Dose} $(I_0 = 5\times10^5)$}  \\ 
\cmidrule(lr){3-4} \cmidrule(lr){5-6}
                                   & \multicolumn{1}{c}{}                                 & \textbf{60 Views}    & \textbf{90 Views}                                         & \textbf{60 Views} & \textbf{90 Views}                                                \\ 
\midrule
\texttt{Analytical}                      & FBP                                                  &  23.25/0.4502           & 26.01/0.5708                                                & 22.92/0.4272        & 25.59/0.5416                                                       \\ 
\midrule
\multirow{2}{*}{\texttt{Supervised}}        & FBPConvNet                                           & 29.67/0.8527 & 30.85/0.8787                                      &  29.53/0.8473        &   30.67/0.8721                                                       \\
                                   & RegFormer                                            &      \underline{33.11}/\textbf{0.9173}       &    \underline{33.77}/\underline{0.9256}                                              &   \underline{32.40}/\underline{0.8865}       &    32.92/0.8901                                                      \\ 
\midrule
\multirow{3}{*}{\texttt{Unsupervised}}      & CoIL                                                 &    30.04/0.8555         &     31.84/0.9001                                            &   30.03/0.8557        &        31.74/0.8966                                                  \\
                                   & SCOPE                                                & 32.00/0.8782             & 33.67/0.9139                                                 &    31.68/0.8657      &  \underline{33.29}/\underline{0.9007}                                                        \\
                                   & Spener (Ours)                                               &   \textbf{34.33}/\underline{0.9170}          &      \textbf{36.61}/\textbf{0.9424}                                            &  \textbf{34.03}/\textbf{0.9110}        &    \textbf{36.17}/\textbf{0.9349}                                                      \\
\bottomrule
\end{tabular}
\caption{Quantitative results of compared methods on AAPM dataset under different dose setting. The best performance is highlighted in \textbf{bold}, and the second best is \underline{underlined}.}
\label{table2}
\end{table*}

To mimic the real-world CT acquisition, we introduce a Poisson noise model ~\cite{elbakri2002statistical} into the sinogram, defined by $\mathbf{Y}(\mathbf{r}) \sim \mathrm{Poisson} \left\{ I_0 \times \mathrm{e}^{-\mathbf{y}(\mathbf{r})} + \epsilon \right\}$, where $\mathbf{Y}(\mathbf{r})$ denotes the transmitted X-ray photon intensity, $I_0$ is the incident X-ray photon intensity and $\epsilon$ denotes the mean of the background events and read-out noise variance. In the experiment, we set $I_0 = 10^6$ as normal dose and $I_0 = 5\times10^5$ as low dose acquisition setting.

\subsubsection{Implementation Details}
In the Spener, we adopt the FBP reconstruction result as the initial prior image. 
The image encoder includes a two-layer CNN network, with a convolution kernel size of 3 and a feature dim of 48. 
For the hash encoding used in our model, we set its hyper-parameters as follows: base resolution $N_\text{min}=2$, maximal hash table size $T=2^{24}$, and resolution growth rate $b=1.95$. 
For the iteration process, the training epochs are configured as follows: 1000 epochs for $t=1$ and 250 for the subsequent iterations. 
for the regularization, we utilize the BM3D denoiser with $\sigma = 0.01$ . 
Due to space constraints, additional hyperparameter settings for network training are provided in the supplementary materials.

\subsubsection{Methods in Comparison \& Metrics}
We compare Spener to five representative methods: (1) one analytical reconstruction algorithm (FBP); (2) two supervised DL models (FBPConvNet~\cite{jin2017deep} and RegFormer~\cite{RegFormer}); and (3) two unsupervised INR-based reconstruction methods (CoIL~\cite{sun2021coil} and SCOPE~\cite{wu2022self}). To ensure a fair comparison, we use the codebase released by the authors. All approaches are trained and tested on the same datasets, as detailed in the datasets section.

\par For evaluating the SVCT reconstruction performance, we utilize two classic metrics, peak signal-to-noise ratio (PSNR) and structural similarity index measure (SSIM), to assess the reconstruction performance.

\subsection{Comparison with SOTA Methods}
\subsubsection{SVCT Reconstructions in Noise-free Scenario}
Table~\ref{table1} shows the quantitative results.   
On the AAPM dataset (in-domain dataset), our Spener achieves comparable performances with the supervised method RegFormer.
Specifically, our Spener produces the best performance in PSNR and the second-best in SSIM. 
However, on the COVID-19 dataset, which features different CT acquisition geometry, we observe that the two supervised methods (FBPConvNet and RegFormer), trained on the AAPM dataset, suffer from performance drops due to the OOD problem. 
In contrast, the unsupervised methods demonstrate flexibility and effectiveness, with our Spener achieving the best performance.

Similarly, on the CMB-CRC dataset, the two supervised methods (FBPConvNet and RegFormer) encounter the OOD problem due to data from different organs, resulting in reduced performance. However, the unsupervised methods, particularly our Spener model, achieve better results.

Figure~\ref{fig1} shows the qualitative results, which align with the quantitative comparisons above. 
On the AAPM dataset, both RegFormer and Spener recover the desired images.
However, on the COVID-19 dataset, Spener produces cleaner and more detailed CT reconstructions.
Specifically, Spener reconstructs the black `holes', which are potentially caused by emphysema, more clearly, as indicated by the dashed circle.
Additionally, Spener provides a clearer reconstruction of small lung tissues, as indicated by the arrows.
On the CMB-CRC dataset, all methods except our Spener exhibit streaking artifacts.
Spener consistently presents cleaner and more detailed CT reconstructions across all datasets.

\subsubsection{SVCT Reconstructions in Normal and Low Dose}
Table~\ref{table2} shows the quantitative results on the AAPM datasets with 60 and 90 input views under both normal and low dose settings. 
Under the normal dose setting, our Spener achieves the best performance in most cases.
Compared to the two supervised methods, Spener also shows notable performance improvements.
Under the low dose setting, our Spener demonstrates even more significant improvements. 
For instance, PSNR respectively improves by 2.84 dB (36.61 vs. 33.77) and 3.25 dB (36.17 vs. 32.92) with 90 input views. 
These results indicate that Spener is more robust under dose reduction conditions, which more closely align with real-world CT acquisition scenarios.

Figure~\ref{fig2} shows the qualitative results. 
As shown in the figure, the current unsupervised INR methods (CoIL and SCOPE) are sensitive to the noise.
Specifically, CoIL fails to produce satisfactory results, losing high-frequency details and exhibiting persistent streaking artifacts and blurring.
SCOPE reconstructions display some inconsistency points and high-frequency artifacts.
However, despite the noise affecting high-frequency details, our proposed Spener successfully recovers these details, as indicated by arrows.
Remarkably, Spener reconstructs clean and fine-detail CT images for all dose settings, demonstrating robustness to noise.

\subsection{Ablation Studies}
\subsubsection{Effectiveness of Iterative Reconsturction}
In the Spener framework, the iterative reconstruction process for updating the prior image is crucial for optimal performance. 
To evaluate the effectiveness of the iteration, we conduct an experiment comparing the performance of Spener with and without iterative prior image updates in SVCT reconstruction.  
In the non-iterative scenario, Spener uses the FBP reconstruction as the prior image throughout the entire network optimization process.

\begin{figure}[!t]
\centering
\includegraphics[width=\columnwidth]{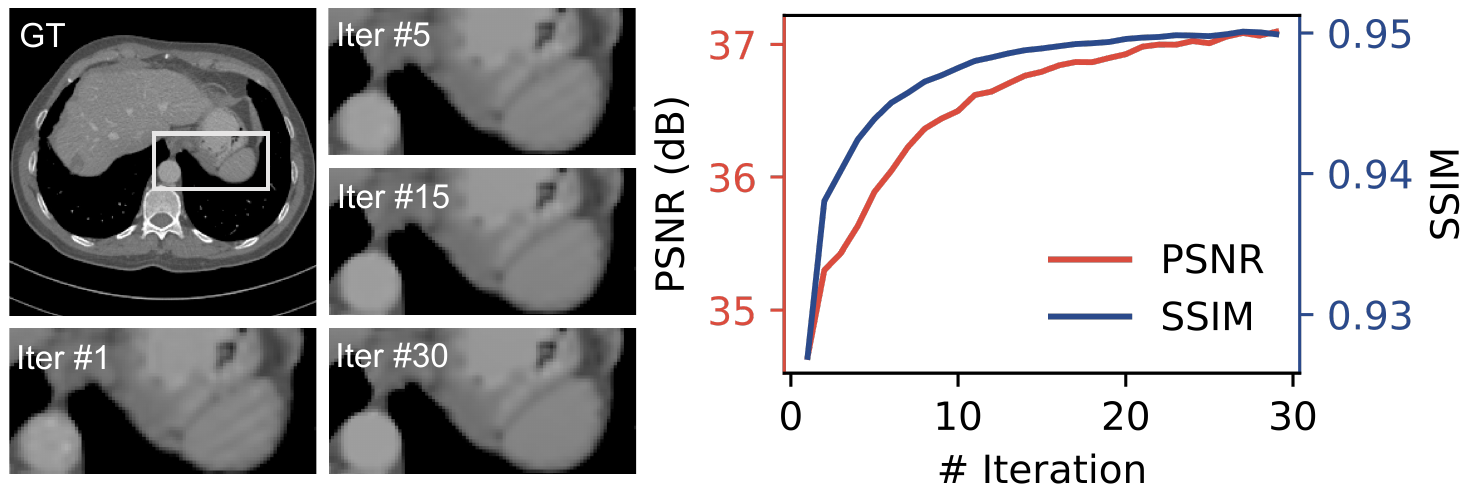} 
\caption{Qualitative and quantitative results of Spener across different iterations on the AAPM dataset with 90 views.}
\label{fig:abl_iteration}
\end{figure}

Table~\ref{tab:abl_iteration} presents the quantitative results. 
With iterative updates, Spener achieves significant performance improvements, with approximately a 1 dB increase in PSNR and a 0.014 gain in SSIM. 
Figure~\ref{fig:abl_iteration} illustrates the qualitative results and performance curves of Spener across different iterations. 
On the left side of the figure, the progressive reduction of streaking artifacts is evident. 
The performance curve indicates that quantitative results consistently improve with each iteration, stabilizing around iteration 20.

\begin{table}
\centering
\begin{tabular}{lcc} 
\toprule
\textbf{Strategy}       & \textbf{PSNR}                            & \textbf{SSIM }              \\ 
\midrule
w/o iteration & 36.28$\pm$0.69 & 0.9342$\pm$0.0060  \\
w/  iteration  & \textbf{37.35$\pm$0.54} & \textbf{0.9484$\pm$0.0062}  \\
\bottomrule
\end{tabular}
\caption{
Quantitative results of CT images reconstructed by Spener with and without iteration on the AAPM dataset with 90 views.}
\label{tab:abl_iteration}
\end{table}

\begin{table}[h]
\centering
\begin{tabular}{lcc} 
\toprule
\textbf{Strategy }    & \textbf{PSNR}           & \textbf{SSIM }              \\ 
\midrule
w/o encoder & 35.26$\pm$0.56 & 0.9381$\pm$0.0068  \\
w/ encoder  & \textbf{37.35$\pm$0.54} & \textbf{0.9484$\pm$0.0062}  \\
\bottomrule
\end{tabular}
\caption{Quantitative results of CT images reconstructed by Spener with and without an image encoder on the AAPM dataset with 90 views.}
\label{tab:abl_encoder}
\end{table}

\subsubsection{Effectiveness of Encoder}
The image encoder in Spener plays a crucial role in extracting effective local image domain priors to enhance network optimization.
To explore the effectiveness of the image encoder in Spener, we conducted a comparisons between Spener with and without the image encoder. 
Specifically, in the absence of an image encoder,  the network directly queries the corresponding image intensity from the prior image for prior information during optimization.
As shown in Table~\ref{tab:abl_encoder}, Spener with image encoder achieves significant improvements compared to the version without the encoder.

\begin{table}[H]
\centering

\begin{tabular}{lccc} 
\toprule
\textbf{Settings}         & \textbf{Denoiser} & \textbf{PSNR} & \textbf{SSIM}  \\ 
\midrule
\multirow{2}{*}{No Noise} & $\times$      & 37.37$\pm$0.32    & 0.9434$\pm$0.0050  \\
                          & $\surd$      & \textbf{37.55$\pm$0.63}    & \textbf{0.9458$\pm$0.0056}  \\ 
\cmidrule(lr){1-4}
\multirow{2}{*}{Low Dose} & $\times$      & 35.17$\pm$0.41    & 0.9179$\pm$0.0065  \\
                          & $\surd$       & \textbf{36.03$\pm$0.61}    & \textbf{0.9309$\pm$0.0051}  \\
\bottomrule
\end{tabular}
\caption{Quantitative comparisons of the impact of denoiser regularization on the AAPM dataset with 90 view reconstruction under noise-free and low-dose acquisition settings.}
\label{tab:abl_denosier}
\end{table}

\subsubsection{Influence of Denoiser Regularization}
To investigate the impact of denoiser regularization in iterative reconstruction, we evaluated Spener with and without a denoiser on AAPM 90-view reconstructions under no-noise and low-dose conditions. As shown in Table~\ref{tab:abl_denosier}, Spener with denoiser outperforms the version without, particularly under low-dose settings, with notable improvements in PSNR and SSIM.

\begin{table}
\centering

\begin{tabular}{lccc} 
\toprule
\textbf{Settings}         &  \textbf{$\lambda$} & \textbf{PSNR} & \textbf{SSIM}  \\ 
\midrule
\multirow{3}{*}{No Noise} & $0.0$      & 37.26$\pm$0.61    & \textbf{0.9456$\pm$0.0056}  \\
                          & $2.5$      & \textbf{37.29$\pm$0.60}    & 0.9449$\pm$0.0062  \\ 
                          & $5.0$      & 37.16$\pm$0.52    & 0.9443$\pm$0.0054  \\ 
\cmidrule(lr){1-4}
\multirow{3}{*}{Low Dose} & $0.0$      & 35.99$\pm$0.44    & 0.9279$\pm$0.0080  \\
                          & $2.5$       & \textbf{36.10$\pm$0.49}    & 0.9321$\pm$0.0043  \\
                          & $5.0$      & 35.95$\pm$0.53    & \textbf{0.9322$\pm$0.0045}  \\ 
\bottomrule
\end{tabular}
\caption{Quantitative comparisons of the influence of $\lambda$ on AAPM dataset with 90 view reconstruction under noise-free and low-dose acquisition settings.}
\label{tab:abl_lambda}
\end{table}

\subsubsection{The Choice of Hyperparameter $\lambda$}
To analyze the influence of $\lambda$ in network optimization, we evaluated the performance of Spener with $\lambda$ sets as 0, 2.5, and 5 on AAPM 90-view reconstruction under both noise-free and low-dose conditions. 
Table~\ref{tab:abl_lambda} shows the quantitative results. 
Overall, the different values of $\lambda$ do not significantly impact the performance of Spener. 
In a noise-free setting, Spener with $\lambda= 0$ and $\lambda= 2.5$ achieve a comparable performance. 
While with a low dose setting, Spener with $\lambda= 2.5$ yields better performance with an approximate 0.1 dB improvement in PSNR.

\begin{figure}[!t]
\centering
\includegraphics[width=\columnwidth]{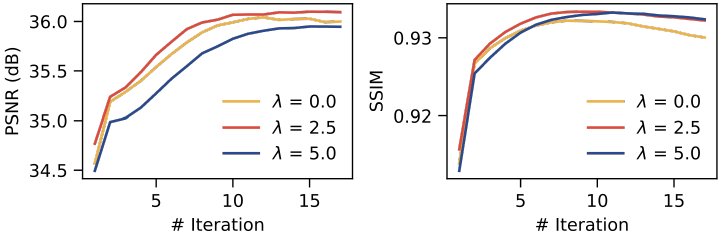} 
\caption{Performance curves of Spener with different $\lambda$ settings on AAPM dataset with 90 views reconstruction under low-dose acquisition.}
\label{fig:abl_iteration_curve}
\end{figure}

Figure~\ref{fig:abl_iteration_curve} illustrates the performance curves with different $\lambda$ settings under low-dose conditions. Spener's performance steadily improves across all $\lambda$ settings during the initial iterations, with $\lambda=2.5$ achieving the best results. 
As iterations continue, a noticeable decline in SSIM occurs when $\lambda =0$ demonstrates the importance of regularization loss in noisy scenarios.
Spener with $\lambda=2.5$ provides the optimal performance. Therefore, we set $\lambda = 2.5$ for the all experiments.

\section{Conclusion \& Discussions}
This work introduces Spener, a novel iterative unsupervised SVCT reconstruction method. Unlike SOTA supervised methods, Spener does not require external training data, making it more practical in clinical scenarios with varying acquisition conditions. Compared to SOTA unsupervised INR-based methods, Spener addresses their performance degradation in highly ill-posed SVCT reconstructions, particularly when the measurement data is noisy. Experimental results demonstrate that Spener outperforms current SOTA methods.
Despite the success achieved with the current framework in SVCT, there are several areas for improvement: 1) exploring alternative iterative frameworks with INR methods to accelerate model convergence; 2) adopting a more effective image domain encoder to provide robust image priors, enhancing both training speed and performance; 3) extending the framework to address other inverse problems in medical imaging.

\section{Acknowledgments}
This work was supported by the National Natural Science Foundation of China under Grant 62071299. 

\bibliography{aaai25}

\end{document}